\documentclass[aps,pre,twocolumn,groupedaddress,showpacs]{revtex4}
\usepackage{graphicx}
\usepackage{amssymb}

\begin{document}
\title{Granular discharge and clogging for tilted hoppers}

\author{Hannah G. Sheldon and D. J. Durian}
\affiliation{Department of Physics \& Astronomy, University of
Pennsylvania, Philadelphia, PA 19104-6396}


\date{\today}

\begin{abstract}
We measure the flux of spherical glass beads through a hole as a systematic function of both tilt angle and hole diameter, for two different size beads.  The discharge increases with hole diameter in accord with the Beverloo relation for both horizontal and vertical holes, but in the latter case with a larger small-hole cutoff.  For large holes the flux decreases linearly in cosine of the tilt angle, vanishing smoothly somewhat below the angle of repose.  For small holes it vanishes abruptly at a smaller angle.  The conditions for zero flux are discussed in the context of a {\it clogging phase diagram} of flow state vs tilt angle and ratio of hole to grain size.
\end{abstract}

\pacs{45.70.-n}


\maketitle


%
%

\section{Introduction}

The flow of granular materials is of widespread practical \cite{Nedderman, Muzzio02} and fundamental \cite{JNBrev96, DuranBook, LiuNagelBOOK} interest.  One challenge to understanding and controlling behavior is that the response is nonlinear, with a forcing threshold below which the medium is static.  Furthermore, just above threshold the response may be intermittent even though the forcing is steady.  Familiar examples include avalanches down the surface of a heap as well as gravity-driven discharge from a horizontal hole at the bottom of a deep container or ``silo''.  For the latter, the mass discharged per unit time is given by the ``Beverloo'' relation:
\begin{equation}
W=C\rho_b g^{1/2}(D-kd)^{5/2},
\label{beverlooequation}
\end{equation}
where $\rho_b$ is the density of the bulk granular medium, $g=980$~cm/s$^2$, $D$ is the hole diameter, $d$ is the grain diameter, and $C$ and $k$ are dimensionless fitting parameters \cite{beverloo}.  By contrast with viscous fluids, the discharge of grains is independent of filling depth. This may be understood in terms of the classic Janssen argument that pressure vs depth approaches a constant for a deep container due to support of the weight of the medium by frictional contacts with the sidewalls.  While the Beverloo relation is supported by a large body of work, as reviewed by Nedderman and Savage et al.~\cite{NeddermanSavage}, discrepancies of up to forty percent have been recently reported when the hole size is increased more widely than usual~\cite{MazaGM07}.  Typical ranges for the numerical constants are $0.5<C<0.7$ and $1.2<k<3$, depending on grain shape and friction.  

The Beverloo equation implies the existence of a threshold hole diameter, $kd$, of a few grains across, below which the flux vanishes.  Just above this threshold, the flow is subject to intermittent clogging~\cite{HerrmannEPJE00, MazaPRE03, MazaPRE05, Franklin09}.  Even far above threshold, the response may not be steady in that the Beverloo form is often interpreted in terms of intermittent formation and breakup of arches across the hole.  In particular, grains in a freshly-broken arch free-fall through a distance set by hole size and hence emerge with speed $v\sim (gD)^{1/2}$ and mass per unit time $W\sim \rho_b g^{1/2}D^{5/2}$.  Density waves \cite{BehringerPRL89, Raafat} and ticking \cite{Schick, xlwPRL93, Veje} are also examples of unsteady response in gravity-driven discharge, but where air plays a role.

\begin{figure}
\includegraphics[width=2.25in]{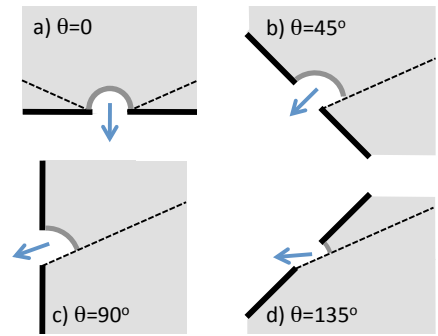}
\caption{Schematic cross-section of the discharge apparatus tilted at various angles, as labeled.  The container wall is indicated by the heavy black lines, with a circular hole through which the grains emerge.  The granular material is indicated by the gray shaded region.  Below some boundary indicated by the thin dashed line the grains may be stagnant.  Transient arches may extend across the hole, as indicated by the thick dark gray curves.
}\label{schematic}
\end{figure}

To develop a deeper microscopic understanding of granular discharge, it seems important to grapple with the unsteadiness represented by intermittent clogging and arch formation / breakup.  One experimental approach is to vibrate the system, in order to fluidize and break arches as well as to introduce a competing time scale \cite{Evesque93, HuntPF99, WassgrenPF02, SchifferPRE06, PachecoPRE08}.  In this paper, our approach is to interfere with the usual arch formation and breakup by tilting the container as depicted in Fig.~\ref{schematic}, so that the plane of the hole is inclined by an angle $\theta$ away from horizontal.  According to the ``free-fall arch'' picture, one might expect the discharge rate to decrease according to the reduced horizontal projection of the hole, $\cos\theta$, and to vanish at a tilt angle less than ninety degrees where the projection falls below a nonzero threshold.

There are few prior experiments on granular flux from holes that are not horizontal, as noted in an article concerning the use of vertical slots for constructing a flow meter \cite{Davies91}.  Ref.~\cite{FranklinCES55} predates Beverloo and reports flow rates for two media and three hole diameters at inclination angles of \{0, 30, 60, 90\} degrees.  The results are claimed to be linear in the cosine of the tilt angle; however, we concur with statements in Ref.~\cite{Davies91} that the data are too sparse and uncertain to demonstrate the proposed form.  The definitive review by Nedderman et al.~\cite{NeddermanSavage} cites Ref.~\cite{Chitty70} as a ``preliminary investigation which comes to no clear conclusion'' regarding discharge through a vertical hole; it also cites Ref.~\cite{FranklinCES55} but only regarding horizontal holes.  Ref.~\cite{Chang91} reports that the discharge rate for a vertical hole at the very bottom of a sidewall scales as diameter to a power between 2.5 and 2.8, and that the ratio of vertical to horizontal discharge rates is between 0.37 and 0.50.  Ref.~\cite{BehringerPRE07} reports on velocity fields, but not discharge rates, for conical hoppers tilted up to $23^\circ$.  Thus, the observations reported here concern a relatively unexplored effect.


\section{Materials and Methods}

The granular medium consists of spherical glass beads, with two different diameters: $d=0.30\pm0.05$~mm and $d=0.9\pm0.1$~mm.  Both have bulk density $\rho_b=1.53\pm0.01$~g/cm$^3$ and draining angle of repose $\theta_r=24^\circ$.  Two different types of container are used.  The first type is a steel can, with 10~cm diameter, 12~cm height, and 0.25~mm wall thickness.  Holes are drilled in three different locations: in the bottom at center, in the bottom at 2~cm from the side, and in the side at 2~cm above the bottom.  The second container type is square Aluminum tubing, with $7\times7$~cm$^2$ inner cross section, 31~cm height, and 3~mm wall thickness.  One to four holes are drilled in each side in a staggered arrangement near the bottom, no closer than 2~cm from any edge or from each other, and are countersunk with a $120^\circ$ chamfer.  The containers are open at the top, so there is no back-flow of air into the hole to replace the loss of granular material.  The containers are grounded to prevent electrostatic charging, and are mounted by a chain clamp with changeable orientation on a ring stand.  The tilt angle $\theta$ of the plane of the hole away from horizontal is measured with a plumb bob and protractor; $\theta=0^\circ$ corresponds to a horizontal hole, $\theta=90^\circ$ corresponds to a vertical hole, and $\theta>90^\circ$ corresponds to discharge with an upward angle of $\theta-90^\circ$; see Fig.~\ref{schematic} for a sketch of the system at four different tilt angles.  Flow is impossible for $\theta>180^\circ-\theta_r=156^\circ$, where the granular medium loses contact with the boundary around the hole.  

Discharge rates are measured by weighing the material collected during a timed interval ranging from several seconds for fast flows, to several minutes for slow flows, as follows.  First the hole is covered with a piece of paper and the granular medium is poured from a beaker into the container, usually to its full height.  Then the covering is removed and flow allowed to proceed for a several seconds or more.  Next a beaker is inserted into the discharge stream while simultaneously starting a timer.  Finally, the beaker is removed from the stream while simultaneously stopping the timer.   Statistical uncertainty in discharge rate is typically one to ten percent, as reflected by the size of the scatter in the data for runs taken under identical or similar conditions; error bars are not displayed since they are smaller or comparable to symbol size.


\section{Discharge Rate}

\begin{figure}
\includegraphics[width=3.000in]{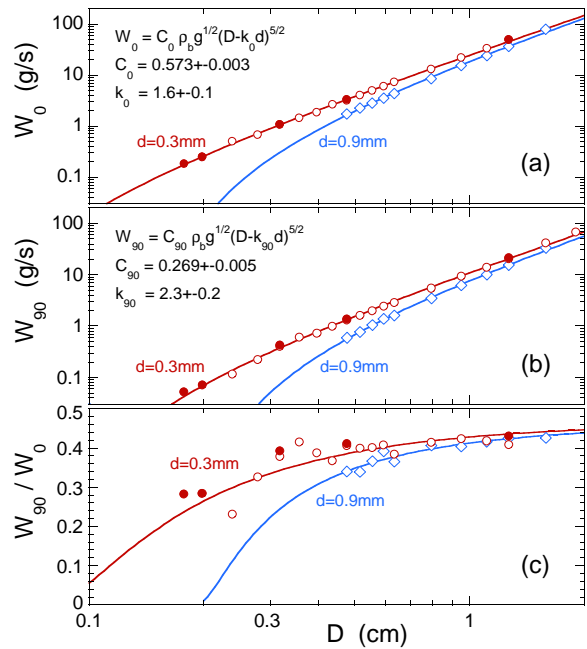}
\caption{(Color online) Discharge rate vs hole diameter for (a) horizontal and (b) vertical holes; the ratio is shown in (c).  Red circles (blue diamonds) are for $d=0.3$~mm (0.9~mm) diameter glass beads.  Open symbols are for containers made from Al tubes of square cross section; solid symbols are for steel cans. Solid curves are two-parameter fits to the Beverloo form, as specified; the ratio of these fits is shown in (c).}\label{beverloo}
\end{figure}

We begin by considering the masses per unit time, $W_0$ and $W_{90}$, respectively discharged through horizontal and vertical holes.   Initial attention is restricted to hole sizes that are large enough not to exhibit clogging or to require tapping to initiate flow.  The first observation is that the discharge rates do not depend on the filling depth, which was varied from the full container height to about one-third of the container height for both the steel cans and the Aluminum tubes.   Care is thus taken to ensure that measurements are not corrupted by any change in discharge rate that may arise as the medium fully empties from the container and the upper free surface comes near the orifice.  This well-know behavior contrasts with liquid discharge, and is explained by the classic Janssen argument. Furthermore, the discharge rates do not depend on details of the container geometry, which was varied in terms of container shape and hole location as described above.  This is also evident in Figs.~\ref{beverloo}(a,b), where discharge rates for the two container types are plotted versus hole diameter, for both horizontal and vertical holes and for both grain sizes.  

The observed discharge rates all increase with hole size, more rapidly for smaller holes, and appear to approach a power-law for larger holes.  Fits of discharge data to the Beverloo relation are included in Figs.~\ref{beverloo}(a,b) as solid curves.  The agreement is good and furthermore the numerical constants are independent of bead size: \{$C_0=0.573\pm0.003$, $k_0=1.6\pm0.1$\} for horizontal holes and \{$C_{90}=0.269\pm0.005$, $k_{90}=2.3\pm0.2$\} for vertical holes.  Note that, therefore, the two curves in each plot represent one simultaneous fit.  The fitting parameters for horizontal holes in Fig.~\ref{beverloo}(a) are in accord with prior work~\cite{NeddermanSavage}.  The fitting parameters for vertical holes in Fig.~\ref{beverloo}(b) indicate both that the zero-flux threshold hole diameter is larger than for horizontal holes, $k_{90}d>k_{0}d$, and that for very large holes the ratio of vertical to horizontal discharge rates becomes constant.  This can be seen directly in the discharge ratio, $W_{90}/W_0$, plotted in Fig.~\ref{beverloo}(c).  The observed ratios increase with hole size and fall in the range $0.3-0.5$ reported in Ref.~\cite{Chang91}.  Note, however, that the asymptotic value $C_{90}/C_0=0.47$ from the Beverloo fits is not attained even for the largest holes studied.

We have no theoretical explanation for the observed diameter-dependence of  the discharge rate for vertical holes, other than that the basic scale must be set dimensionally as $\rho_bg^{1/2}D^{5/2}$ with a numerical prefactor that depends on inclination.  The fact that the Beverloo form successfully describes both horizontal and vertical discharge rates raises questions about the physical interpretation in terms of free-fall through a distance set by hole size, since the component of grain velocity that contributes to discharge is perpendicular to gravity for vertical holes.   Transient arches could extend from the top of the hole to the solid packing of stagnant grains below a boundary, perhaps set by the angle of repose, as sketched in Fig.~\ref{schematic}; however, upon breaking, the grains in such an arch cannot escape at free-fall speed.  Any free-fall will be terminated by collision with either the stagnant pile or with other grains in a flow field that curves toward the exit.  It would be interesting to map out the flow field and the time-averaged density for tilted hoppers, even in a quasi-two dimensional system as in~Refs.\cite{NeddermanSavage2, HuntPF99, WassgrenPF02, Kudrolli05}, to directly investigate such behavior.   In addition, a further question is raised for how to interpret the different zero-flux threshold hole diameters, $k_{0}d \ne k_{90}d$, in terms of an ``empty annulus'' where grain centers may not pass \cite{NeddermanSavage} because that concept is independent of hole orientation.

\begin{figure}
\includegraphics[width=3.000in]{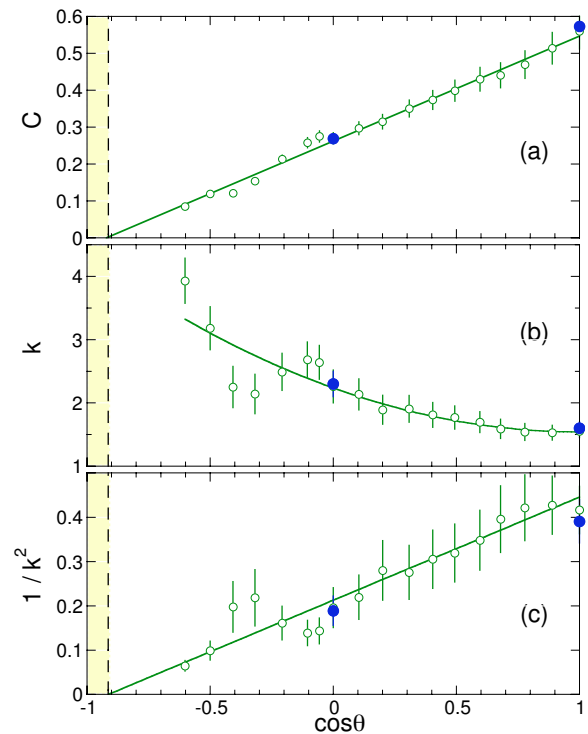}
\caption{(Color online)  Dimensionless parameters, $C$ and $k$, obtained by fitting flux vs diameter data to the Beverloo relation, $W = C \rho_b g^{1/2}(D-kd)^{5/2}$, for holes oriented at angle $\theta$ away from horizontal.  Solid symbols are for the fits shown in Fig.~\protect{\ref{beverloo}}, and open symbols are from fits to data shown in Figs.~\protect{\ref{flux300}}-\protect{\ref{flux900}}. Error bars are based on both uncertainty in fits as well as on the difference of results for the two grain sizes.  The vertical dashed line and shaded region indicate $\cos\theta<-\cos\theta_r$ where discharge is impossible.  The solid lines in (a) and (c) are the best fits to a line; the solid curve in (b) is the best fit to a parabola with minimum at $\cos\theta=1$.} \label{bevfits}
\end{figure}

Since there is no theory to test, we now examine trends empirically two different ways.  First, flux data are obtained for a wide range of tilt angles, not just zero and ninety degrees, for both size grains and for three to six different hole diameters.  We find that the Beverloo equation gives satisfactory fits to flux vs diameter in all cases.  Results for the fitting parameters, $C$ and $k$, are plotted vs cosine of the tilt angle in Fig.~\ref{bevfits}.   The values displayed are an average for both $d=300~\mu$m and $d=900~\mu$m diameter grains.    Note in the top plot that $C$ decreases with tilt angle and appears to be a linear function of $\cos\theta$.  Furthermore, it extrapolates to zero at $-\cos\theta_r$, below which no flow is possible.  Note in the middle plot that $k$ increases with tilt angle, though the functional form is not as clear.  It appears to depart quadratically from the zero-angle value.  And since $C$ apparently vanishes at $\theta_r$, it is natural to speculate that  $k$ diverges at the same angle.  This possibility is reinforced in the bottom plot, in which $1/k^2$ appears to be a linear function of $\cos\theta$ and to vanish at $-\cos\theta_r$.

\begin{figure}
\includegraphics[width=3.000in]{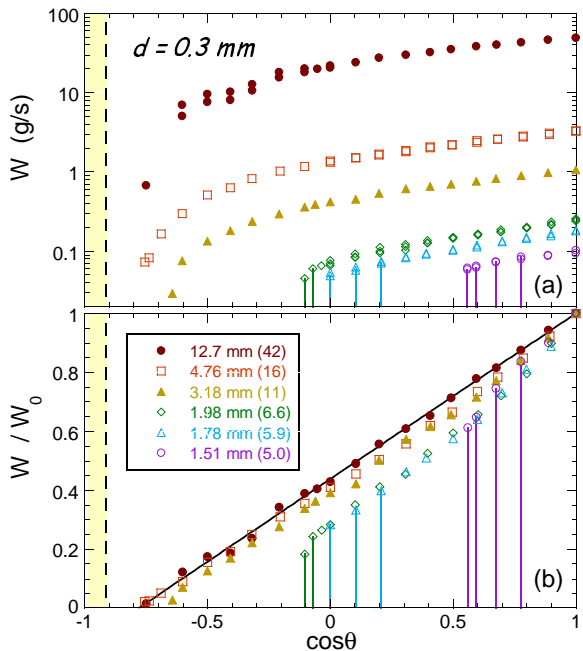}
\caption{(Color online)
(a) Discharge rate vs cosine of tilt angle for $d=0.3$~mm diameter beads through various hole diameters $D$, as labeled; values of $D/d$ are given in parenthesis.  (b) Discharge rate normalized by value at zero tilt angle, with same symbol codes.  The vertical dashed line and shaded region indicate $\cos\theta<-\cos\theta_r$ where discharge is impossible.  The data points with stem lines indicate angles at which the flow is subject to clogging.  In (a) multiple points at a given angle represent data taken with different container types and hole locations; in (b) each point represents the average over different container/hole geometries, and the black line represents the best linear fit to the large-hole data.
}\label{flux300}
\end{figure}
\begin{figure}
\includegraphics[width=3.000in]{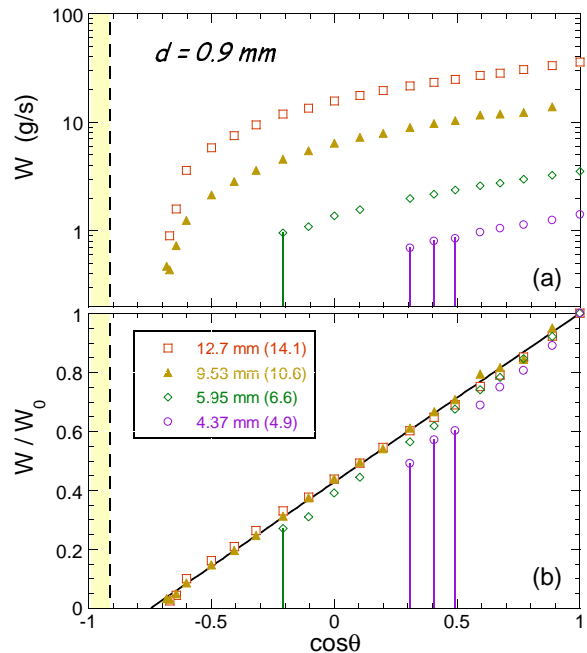}
\caption{(Color online)
(a) Discharge rate vs cosine of tilt angle for $d=0.9$~mm diameter beads through various hole diameters $D$, as labeled, in the square Al tube containers; values of $D/d$ are given in parenthesis.  (b) Discharge rate normalized by value at zero tilt angle, with same symbol codes.  The vertical dashed line and shaded region indicate the region $\cos\theta<-\cos\theta_r$ where discharge is impossible.  The data points with stem lines indicate angles at which the flow is subject to clogging.  The black line in (b) represents the best linear fit to the large-hole data.
}\label{flux900}
\end{figure}

Alternatively, the tilt-dependence of the flux can be considered empirically without using the Beverloo relation.  Specifically, for a given hole size, we examine how tilting causes the flux to decrease from a maximum at zero-angle, $W_0$.  The simplest hypothesis would be reduction according to the projected horizontal area of the hole, i.e.\ $W=W_0\cos\theta$.  So we plot discharge rates for various hole diameters vs $\cos\theta$ in Fig.~\ref{flux300} for $d=0.3$~mm diameter grains.  Raw data are shown on a logarithmic scale in part (a), while the ratio $W/W_0$ of flux to the value at $\theta=0$ is shown on a linear scale in part (b).  There are several interesting features in these plots.  First, as shown by multiple data points at a given angle, the results are independent of container geometry: square tubes and cylindrical cans with different hole placements all have the same discharge rates.  Second, the data sets appear nearly parallel on the logarithmic scale of part (a), and hence are nearly proportional.  However, the data sets are not truly proportional to one another since good collapse is not found in part (b).  For large holes, though, the data appear to approach a common linear dependence on cosine of tilt angle: $W/W_0=(\cos\theta+\alpha)/(1+\alpha)$ where the fitting parameter is $\alpha=0.78\pm0.01$.  We stress that this form is linear in, but not proportional to, $\cos\theta$; therefore, the flux is not proportional to the horizontal projected area.  Moreover, there is a nonzero flux even for $\theta>90^\circ$, when the hole is tilted past vertical and the unit normal vector to the hole has an {\it upwards} component.  Naturally no flow is possible for $\theta>180^\circ-\theta_r=156^\circ$, when the medium loses contact with the boundary surrounding the hole.  This argument underlies the speculation in Ref.~\cite{FranklinCES55} that the value of $\alpha$ should be $\cos\theta_r$. In fact, the flux vanishes below this bound at an angle $\arccos(-0.78)=140^\circ$.  All these features, including the value of $\alpha$, can also be seen in Fig.~\ref{flux900} for $d=0.9$~mm grains.  Theoretical guidance would be helpful in determining whether the best description of discharge rates should be formulated in terms of Beverloo fit parameters, as in Fig.~\ref{bevfits}, or in terms of reduction from zero-tilt discharge, as in Figs.~\ref{flux300}-\ref{flux900}.

We note that the Beverloo parameters in Fig.~\ref{bevfits} were obtained by fits to the raw data displayed in the top plots of Figs.~\ref{flux300} and \ref{flux900}, reorganized as a function of diameter at a given angle.


\section{Clogging}

In this final section we consider whether or not the system is clogged or flowing, and how the state of behavior depends on hole size and tilt angle.  One might suppose that the zero-flux threshold hole diameter, $kd$, found from fits of the Beverloo form to flux data at different tilt angles would demarcate the phase boundary between clogged and flowing; however, this is not the case.   To investigate, we first note that some features of clogging are already displayed in the plots of $W$ and $W/W_0$ vs tilt angle in Figs.~\ref{flux300}-\ref{flux900}. Specifically, stem lines extending from the data points down to zero are used to indicate conditions where the system was observed to clog. In these cases, smooth steady discharge proceeds for some extended time interval long enough to measure flux, but then suddenly and unpredictably stops.  With gentle tapping or poking, the clog can be broken and an interval of steady flow can be restarted.  Visual inspection reveals that clogging is not caused by impurities or larger grains in the tail of the size distribution that block the hole.  As demonstrated in Figs.~\ref{flux300}-\ref{flux900}, we find no such intermittent clogging for sufficiently large holes; instead, with increasing tilt angle, the discharge rate decreases continuously toward zero.  For smaller holes, intermittent clogging happens at nonzero flux over a range of tilt angles.   Thus the transition between zero and nonzero steady flux is discontinuous, but not necessarily sharp.

\begin{figure}
\includegraphics[width=3.000in]{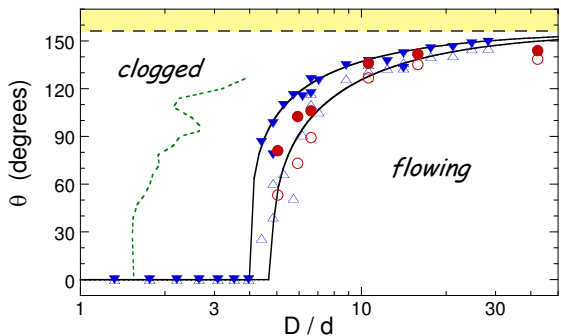}
\caption{(Color online)
Clogging phase diagram of discharge state vs tilt angle and ratio of hole to grain diameter.  Circles are for $d=0.3$~mm grains; triangles are for $d=0.9$~mm grains; open symbols are for stop angles at which flow ceases as the tilt angle is increased; solid symbols are for start angles at which flow commences as the tilt angle is decreased.  The horizontal dashed line and shaded region indicate the region $\theta>180^\circ-\theta_r=156^\circ$ where discharge is impossible.  The solid black curves are guides to the eye that rise abruptly from zero, bracket the transition, and approach $\theta_r$ for large holes.   The dotted green curve represents the normalized zero-flux hole diameter, $k$, from fits to the Beverloo relation, shown previously in Fig.~\protect{\ref{bevfits}}(b).
}\label{clog}
\end{figure}

As a first attempt to locate the transition and ascertain its sharpness we perform the following experiments.   For a given hole size, first we begin at a small tilt angle in the flowing regime and then slowly increase the tilt angle until a clog first occurs; we repeat 2-5 times and record the average ``stop'' angle.  Next we begin at a large tilt angle in the clogged regime and then slowly decrease the tilt angle until until flow commences; we repeat 2-5 times and record the average ``start'' angle.   Tilting is done by hand at a constant rate of roughly one degree per second; this is slow enough, and the motion is smooth enough, that acceleration and jerk do not influence behavior.   Given these steady tilt rates, and the absence of a dwell time at any particular angle, it takes less than a minute to go from $0^\circ$ to stop or from $180^\circ$ to start.   Note that intermittent flow with a lifetime of several minutes or more will appear continuous.   While the stop and start angles may thus depend on tilt rate, the average ought to be less sensitive; this concurs with informal trials at slightly different speeds and with actual data where the tilt rate is only imperfectly controlled by hand.  Thus we construct a ``clogging'' phase diagram in Fig.~\ref{clog} by plotting the stop and start angle data as a function of dimensionless hole diameter $D/d$.  For very small holes where no flow occurs without tapping, we plot the stop and start angles at zero.  For holes larger than about $D/d=4$ the transition angles rise abruptly from zero.  They increase monotonically and appear to approach $180^\circ-\theta_r=156^\circ$ for very large holes.  Note that the stop angle data for different size grains exhibit good collapse for all hole diameters.  By contrast the start angle data appear to collapse well only for large holes.

The sharpness of the transition between clogged and flowing states may be gauged from Fig.~\ref{clog} by the difference in start and stop angles.  The spread in these angles is greatest for small holes, just above $D/d=4$, and decreases steadily for larger holes.  We speculate that the increasing sharpness of the transition may be connected with the simulation result that the distribution of packing densities at which a granular system jams becomes narrower as the system size increases~\cite{OHern03}.  Here, for clogging to occur, the sample need not be jammed everywhere -- only in a volume $V_{\rm hole}\sim D^3$ over the outlet.   Flow proceeds steadily only until a grain configuration arises in $V_{\rm hole}$ that is jammed.  For smaller holes, more packing fractions exist that can jam and the system is more susceptible to clogging.   For large enough holes, no such configurations exists and the system flows freely or not at all.  It would be interesting to explore this possible connection to Ref.~\cite{OHern03} with real-time measurements of packing density in $V_{\rm hole}$.  It would also be interesting to map out the clogged/flowing phase boundary by alternative means, such as contour plots of average flow duration and mass discharged \cite{Charles}.

Lastly we note that the shape of the clogged/flowing phase boundary may be described by empirical curves of the form $\theta = 156^\circ (1-Ad/D)^\alpha$ for $D>Ad$.  The upper solid curve in Fig.~\ref{clog} through the start angles corresponds to $A=4.0$ and $\alpha=1/4$; the lower curve through the stop angles corresponds to $A=4.8$ and $\alpha=1/3$.  These approach $180-\theta_r$ for very large holes; however, we caution that the actual data do not reach this limit and could well saturate at a smaller angle.  Inversion gives the hole diameter for crossover from clogged to flowing as $D = Ad / [1-(\theta/156^\circ)^{1/\alpha}]$. This is quite different from the zero-flux threshold hole diameter $kd$ deduced from fits to the Beverloo equation, shown previously in Fig.~\ref{bevfits}b and now plotted on the clogging phase diagram as a dashed curve.  Comparison shows that that the clogged/flowing hole diameter is about three times larger than the Beverloo zero-flux threshold hole diameter.  Thus the susceptibility to clogging does not directly correspond to the Beverloo equation and the vanishing of flux.  It would be interesting to obtain discharge data for holes below clogging, as the flow duration and flux both approach zero, to see if deviation from the Beverloo form can be detected.

\section{Conclusion}

In this paper we reported on discharge rates and clogging behavior for glass beads and circular apertures as a function of both hole size and inclination angle.  Extensive and systematic variation of the latter serve to fill a particularly unexplored void in the literature.  Our discharge results shed new light on the Beverloo relation, particularly the free-fall arch and empty-annulus interpretations, as well as its validity for small holes.   Our clogging results emphasize the need for theoretical consideration of fluctuation and jamming effects, especially for slow flows.  Altogether, our experiments round out the phenomenology of granular discharge, highlight the unusual and elusive mechanics of granular materials, and suggest specific further lines of research.

\begin{acknowledgments}
We thank T.~Brzinski for experimental assistance, and  S.\ R.~Nagel for an inspiring series of experiments and simulations on granular media and jamming.  Our work was supported by the NSF through grant DMR-0704147 and by the University of Pennsylvania through its work-study financial aid program for undergraduate students.
\end{acknowledgments}


\begin{thebibliography}{99}



\bibitem{Nedderman} R. M. Nedderman, ``Statics and kinematics of granular materials'' (Cambridge University, NY, 1992).

\bibitem{Muzzio02} F. J. Muzzio, T. Shinbrot, and B. J. Glasser, Powder Technology {\bf 124}, 1 (2002).

\bibitem{JNBrev96} H. M. Jaeger, S. R. Nagel, and R. P. Behringer, Rev. Mod. Phys. {\bf 68}, 1259 (1996).

\bibitem{DuranBook} J. Duran, ``Sands, powders, \& grains:~An introduction to the physics of granular materials'' (Springer, NY, 2000).

\bibitem{LiuNagelBOOK} ``Jamming and Rheology: Constrained Dynamics on Microscopic and Macroscopic Scales'' edited by A. J. Liu and S. R. Nagel,  (Taylor \& Francis, NY 2001).





\bibitem{beverloo} W. A. Beverloo, H. A. Leniger, and J. Van de Velde, Chem.
Eng. Sci. {\bf 15}, 260 (1961).

\bibitem{NeddermanSavage} R. M. Nedderman, U. T\"uz\"un, S. B. Savage, and G. T. Houlsby, Chem. Eng. Sci. {\bf 37}, 1597 (1982).

\bibitem{MazaGM07} C. Mankoc, A. Janda, R. Arevalo, J.M. Pastor, I. Zuriguel, A. Garcimartin, and D. Maza, Gran. Matt. {\bf 9}, 407 (2007).




\bibitem{HerrmannEPJE00} S.S. Manna and H.J. Herrmann, Eur. Phys. J. E {\bf 1}, 341 (2000)

\bibitem{MazaPRE03} I. Zuriguel, L. A. Pugnaloni, A. Garcimartin, and D. Maza, Phys. Rev. E {\bf 68}, 030301R (2003).

\bibitem{MazaPRE05} I. Zuriguel, A. Garcimartin, D. Maza, L. A. Pugnaloni, and J. M. Pastor, Phys. Rev. E {\bf 71}, 051303 (2005).

\bibitem{Franklin09} S. Saraf and S. V. Franklin, Bull. Am. Phys. Soc. {\bf 54}, J14.00005 (2009).





\bibitem{BehringerPRL89} G.W. Baxter, R.P. Behringer, T. Fagert, and G.A. Johnson, Phys. Rev. Lett. {\bf 62}, 2825 (1989).

\bibitem{Raafat} T. Raafat, J. P. Hulin, and H. J. Herrmann, Phys. Rev. E {\bf 53}, 4345 (1996).

\bibitem{Schick} K. L. Schick, and A. A. Verveen, Nature {\bf 251}, 599 (1974).

\bibitem{xlwPRL93} X.-l. Wu, K. J. Maloy, A. Hansen, M. Ammi, and D. Bideau, Phys. Rev. Lett. {\bf 71}, 1363 (1993).

\bibitem{Veje} C. T. Veje, and P. Dimon, Phys. Rev. E {\bf 56}, 4376 (1997).





\bibitem{Evesque93} P. Evesque and W. Meftah, Int. J. Mod. Phys. B {\bf 7}, 1799 (1993).

\bibitem{HuntPF99} M. L. Hunt, R. C. Weathers, A. T. Lee, C. E. Brennen, and C. R. Wassgren, Phys. Fluids {\bf 11} 68 (1999).

\bibitem{WassgrenPF02} C. R. Wassgren, M. L. Hunt, P. J. Freese, J. Palamara, and C. E. Brennen, Phys. Fluids {\bf 14} 3439 (2002).

\bibitem{SchifferPRE06} K. Chen, M. B. Stone, R. Barry, M. Lohr, W. McConville, K. Klein, B. L. Sheu, A. J. Morss, T. Scheidemantel, and P. Schiffer, Phys. Rev. E {\bf 74}, 011306 (2006).

\bibitem{PachecoPRE08} H. Pacheco-Martinez, H.J. van Gerner, and J. C. Ruiz-Su\'arez, Phys. Rev E {\bf 77} 021303 (2008).




\bibitem{Davies91} C. E. Davies and J. Foye, Trans. Inst. Chem. Eng. {\bf 69}, 269 (1991).

\bibitem{FranklinCES55} F. C. Franklin and L. N. Johanson, Chem. Eng. Sci. {\bf 4}, 119 (1955).

\bibitem{Chitty70} C. D. Chitty and M. A. Spencer, Chemical Engineering, Tripos Part 2. Research Project Report, University of Cambridge (1970).  Nedderman informs us that this document has been discarded without regret (personal communication).

\bibitem{Chang91} C. S. Chang, H. H. Converse, and J. L. Steele, Trans. Am. Soc. Agr. Eng. {\bf 34}, 1789 (1991).

\bibitem{BehringerPRE07} J. F. Wambaugh, R. P. Behringer, J. V. Matthews, and P. A. Gremaud, Phys. Rev. E {\bf 76}, 051303 (2007)



\bibitem{NeddermanSavage2} U. T\"uz\"un, G. T. Houlsby, R. M. Nedderman, and S. B. Savage, Chem. Eng. Sci. {\bf 37}, 1691 (1982).

\bibitem{Kudrolli05} J. Choi, A. Kudrolli, and M. Z. Bazant, J. Phys.-Cond. Matt. {\bf 17}, S2533 (2005).


\bibitem{OHern03} C. S. O'Hern, L. E. Silbert, A. J. Liu, and S. R. Nagel, Phys. Rev. E {\bf 68}, 011306 (2003).

\bibitem{Charles} C. C. Thomas and D. J. Durian, now in progress.


\end{thebibliography}

\end{document}